# Direct Numerical Simulation of Low and Unitary Prandtl Number Fluids in Reactor Downcomer Geometry

Cheng-Kai Tai[1], Tri Nguyen[2], Arsen S. Iskhakov[1], Elia Merzari[2],
Nam Dinh[1], and Igor A. Bolotnov[1]

[1]Department of Nuclear Engineering, North Carolina State University
Campus Box 7909, Raleigh, NC, USA 27695-7909
ctai2@ncsu.edu, aiskhak@ncsu.edu, ntdinh@ncsu.edu, igor_bolotnov@ncsu.edu

[2]Ken and Mary Alice Lindquist Department of Nuclear Engineering, Pennsylvania State University
206 Hallowell Building, University Park, PA, USA 16802-4400
nguyen.tri@psu.edu; ebm5351@psu.edu

**ABSTRACT**

Buoyancy effect on low-flow condition convective heat transfer of non-conventional coolants, such as liquid metal and molten salts, is a crucial safety factor to advanced reactors under transient or accidental scenarios. The distinct heat transfer characteristics of non-unitary Prandtl fluids and the inherent complexity of the low-flow mixed convection phenomena requires the development of novel turbulent and heat transfer models that are adaptive to different spatiotemporal scales involved in the mixed convection heat transfer. In this work, direct numerical simulation of low-flow mixed convection is carried out at low-to-unitary Prandtl numbers that are of industrial interest. Time-averaged statistics, turbulent Prandtl number, as well as time signals are analyzed to investigate mixed convection phenomenon. From the time-averaged statistics, buoyant plume altered velocity boundary layer as well as the intensity of the fluctuation near both walls and channel centerline. Buoyancy effect also rendered different degree of convective heat transfer enhancement and impairment depends on Prandtl and Richardson number. Analysis of time series was conducted on the sodium mixed convection case to emphasize on the low-Pr mixed convection behavior at transition region. Resulting power spectra density and wavelet spectrogram suggests possible large convective structure in transition region. Future work will focus on providing broader data coverage on $Pr - Re - Ri$ parameter space to facilitate more comprehensive analysis of mixed convection.

## 1. INTRODUCTION

Liquid metal and molten salts are the potential candidates for the cooling media in the primary side of the advanced nuclear reactor designs [1]. In contrast to the light water that is widely adopted in current commercial reactors, liquid metal and molten salts are categorized as low- and high-Prandtl (low-Pr and high Pr) coolants due to their characteristics in thermal diffusion. Hence, distinction in their heat transfer behaviors pose challenge to the existing turbulent and heat transfer modeling measures for light water reactors. With the comprehensive understanding thermal-hydraulic behavior of non-conventional coolants is not yet established, research campaign in Integrated Research Project (IRP) of Nuclear Energy Advanced Modeling and Simulation (NEAMS) program for thermal-fluid application in nuclear energy aims to identify and gain understanding in heat transfer challenges for advanced reactor-relevant flow conditions and to support the development of modeling techniques [2].

Mixed convection (buoyancy-affected convection) heat transfer at low-flow transients and accident scenarios plays a crucial role to passive safety of next-generation reactors. The current knowledge basis of low-to-unitary Pr mixed convection is constructed from early experimental work of vertical mixed convection of unitary and low-Pr fluids vertical tubes [2, 3, 4, 5], from which effect of buoyancy on convective heat transfer (Nusselt number) is correlated for different Pr of interest. With growth of computational capability in recent years, direct numerical simulation (DNS) and large eddy simulation (LES) studies shed lights on local flow physics of prototypic Poiseuille-Rayleigh-Bénard flow [6, 7], and reactor-relevant geometries such as backward facing step [8] as well as triangular lattice subchannels [9]. On the other hand, high-fidelity mixed convection data for low-Pr fluids is rather scarce and hence hampered development/validation of advanced turbulence and heat transfer models [10, 11]. To bridge the knowledge gap, direct numerical simulation (DNS) of low-flow mixed convection is carried out at Prandtl numbers of liquid metal and molten salts in downcomer-representing canonical flows. Analysis on time signal of velocity and temperature and time-averaged statistics at different orders is discussed in this paper.

The rest of the paper is organized as following: Section 2 gives a detailed description of the problem setup and numerical methods of the performed DNS. Results of the simulation and discussion are provided in Section 3. Section 4 summarizes this work.

## 2. METHODOLOGY

### 2.1. Problem of Interest

Figure 1 shows the domain of interest with its dimensions in Table I and specification of velocity and temperature boundary conditions (BCs) in Table II. The DNS of low-Pr mixed convection is carried out in a vertical planar channel consisting of three sections: the inlet adiabatic, with heat flux (main interest for analysis), and the outlet adiabatic. The flow on the $+z$ direction is driven by velocity inlet boundary condition with velocity profile recycled from the cross section at $z = 5$ to create fully developed turbulence when flow enters the heat flux section. Gravity and heat flux (simultaneous heating and cooling) are applied in the range of $z \in [5,55]$.

### 2.2. Governing Equations and Numerical Methods

In this study, open-source Navier-Stokes solver NekRS, the GPU-oriented variant of Nek5000 [12] is employed [13, 14, 15]. Nek5000 is based on the spectral element method (SEM), which can provide high-order numerical accuracy while keeping low numerical dispersion and diffusion. In addition, the code is massively parallel and is well-suited for large-scale LES and DNS. Inheriting features of Nek5000, NekRS can further leverage GPU architectures to achieve computational speedup over its predecessor. This gives NekRS advantage in the next generation extreme-scale high-performance computing systems, which are GPU-accelerated. Please refer to [16] for detailed introduction to Nek5000/NekRS.

With SEM, computational domain of interest is discretized into $N_{ele}$ boundary-conforming hexahedral elements. Within each spectral elements, the solution variables are expressed as tensor product of Lagrange polynomial of order $N$ based on the Gauss-Lobatto-Legendre (GLL) quadrature points. As a result, there are $N_{ele}(N + 1)^3$ number of degree-of-freedom (DOF) in a problem.

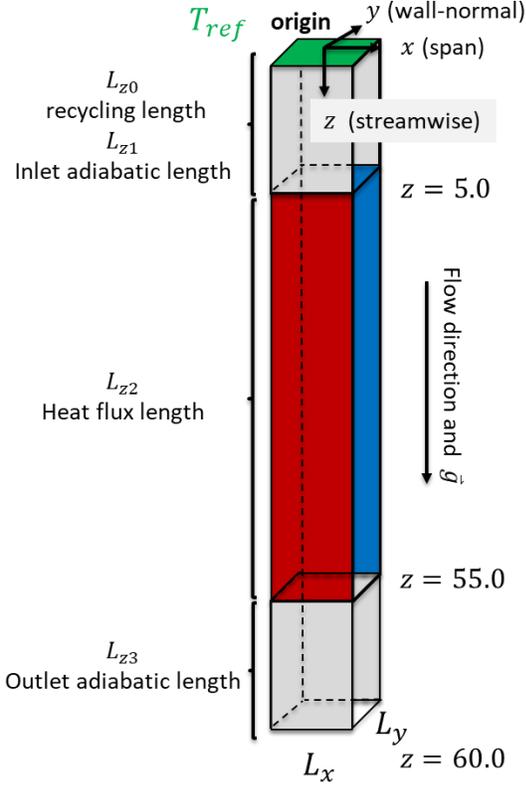

**Figure 1.** Schematic of domain of interest.

**Table I. Geometric specifications.**

| Distance between walls $L_y$ | $L_y = 2\delta$ $\delta = 0.034 m$ |
|---|---|
| Upper/lower adiabatic length ($L_{z0}/L_{z3}$) and recycling length ($L_{z1}$) $L_{z0} = L_{z1} = L_{z3}$ | $L_{z0} = 12\delta$ $L_{z1} = 8\delta$ $L_{z3} = 20\delta$ |
| Heat flux length ($L_{z2}$) | $L_{z2} = 200\delta$ |
| Spanwise length ($L_x$) | $L_x = 2\pi\delta$ |

**Table II. Specification of BCs.**

| | Velocity | Temperature |
|---|---|---|
| $-z$ ($z = 0$) | Velocity inlet (recycling) | Dirichlet |
| $+z$ ($z = 60$) | Stabilized outlet (Dong et al.) | Neumann (zero-gradient) |
| $-y$ ($y = -0.25$) | No-slip wall | Neumann (positive heat flux) |
| $+y$ ($y = 0.25$) | No-slip wall | Neumann (negative heat flux) |
| $\pm x$ | periodic | periodic |

Non-dimensional incompressible Navier-Stokes equations are solved:

$$\nabla \cdot \boldsymbol{u}^* = 0 \tag{1}$$

$$\frac{\partial \boldsymbol{u}^*}{\partial t^*} + \boldsymbol{u}^* \cdot \boldsymbol{\nabla} \boldsymbol{u}^* = -\boldsymbol{\nabla} P^* + \boldsymbol{\nabla} \cdot \frac{1}{Re} \boldsymbol{\nabla} \boldsymbol{u}^* + Ri_q\, T^* \tag{2}$$

$$\frac{\partial T^*}{\partial t^*} + \boldsymbol{u}^* \cdot \boldsymbol{\nabla} T^* = \frac{1}{Pe} \nabla \cdot \nabla T^* \tag{3}$$

$$Ri_q = \frac{Gr_q}{Re^2} \tag{4}$$

$$T^* = \frac{T - T_{ref}}{\Delta T_{ref}} \tag{5}$$

where Eqs (1)-(3) are continuity, momentum, and energy equations, respectively. Boussinesq approximation is applied to account for the temperature feedback to the momentum equation due to the buoyancy effect using Richardson number, Eq. (4), and non-dimensional temperature, Eq. (5). Note that the superscript $*$ is used here to denote dimensionless quantities. The reference scales for non-dimensionalization are listed in Table III. Due to the heat flux BCs, temperature in the domain of interest is part of the solution, *i.e.*, not known *a priori*. Therefore, modified definition for Grashof number is needed in the context of this paper, as shown in (6):

$$Gr_q = \frac{g\beta \Delta T_{ref} L_{ref}^3}{\nu^2} = \frac{g\beta q''_{wall} L_{ref}^4}{2\lambda_{ref} \nu^2} \tag{6}$$

$$\lambda_{ref} = (\rho c_p)_{ref} \alpha_{ref} \tag{7}$$

$$q''_{wall} = \lambda_{ref} \frac{\Delta T_{ref}}{\frac{L_{ref}}{2}} \tag{8}$$

where reference temperature difference (based on pure conduction condition) is used in place of the temperature difference between hot and cold walls in commonly used definition of $Gr$. Also note that the conduction-based temperature is also assumed to be the largest temperature difference possible in the scope of the problem of interest (zero convective heat transfer). Dimensionless wall heat flux applied in the simulation can be derived based on Eq. (8):

$$q''^*_{wall} = \frac{q''_{wall}}{q''_{wall,ref}} = \frac{2}{Pe} = \frac{2}{RePr} \tag{9}$$

**Table III. Reference scales for non-dimensionalization.**

| | |
|---|---|
| Reference velocity $u_{ref}(\frac{m}{s})$ | Bulk velocity ($u_{bulk}$) |
| Reference length scale $L_{ref}(m)$ | Hydraulic diameter ($D_h$) |
| Reference time scale $t_{ref}(s)$ | $D_h/u_{bulk}$ |
| Reference temperature $T_{ref}(K)$ | Inlet temperature ($T_{in}$) |
| Reference density $\rho_{ref}(kg/m^3)$ | $\rho(T_{in}, 1\ atm)$ |
| Reference pressure $p_{ref}\ (Pa)$ | $\rho_{ref} u_{ref}^2$ |
| Reference temperature difference $\Delta T_{ref}\ (K)$ | Conduction temperature difference limit ($\frac{q''_{wall} L_{ref}}{2 k_{ref}}$) |
| Reference heat capacity $(\rho c_p)_{ref}\ (J/K)$ | $\rho_{ref} c_p$ |
| Reference scale of diffusivity $\nu_{ref}, \alpha_{ref}$ ($m^2/s$) | $\nu Re$ |
| Reference wall heat flux $q''_{wall,ref}$ | $(\rho c_p)_{ref} u_{ref} \Delta T_{ref}$ |

## 2.3. Mesh Specifications and Simulation Matrix

To perform mixed convection DNS, both Kolmogorov and Batchelor micro scales have to be resolved for velocity and temperature respectively: $\eta = (\nu^3/\epsilon)^{1/4}$ and $\tau_\eta = (\nu/\epsilon)^{1/2}$. For low- and unitary Pr fluids, Kolmogorov micro scales are the resolution target due to the relatively higher thermal diffusivity. Target resolution (wall, channel centerline, spanwise, streamwise) $(\Delta y_w^+, \Delta y_c^+, \Delta x^+, \Delta z^+)$ is (0.6,6.0,6.0,15). Table IV provide specification of the simulation matrix and mesh.

## 2.4. Data Collection

Catalog of data collected from the simulation matrix is shown in Table V. Time series and different orders of time-averaged statistical moments are collected and discussed in Section 3. Note that the time-averaged statistics are upon whole-domain access so that the physics at whole domain can be examined. Capability of time-averaged statistics collection is verified against the DNS data by Kasagi et al. [17] (noted as reference data set later). The reference case is a turbulent channel with $Re_\tau = 150$ and Dirichlet temperature boundary conditions applied to wall. A benchmark DNS case, summarized in Table VI, is carried out to collect time-averaged statistics and compared against the reference case.

**Table IV. Simulation matrix and mesh specifications.**

| $Re$ | $Pr$ | $Gr_q$ | $Ri_q$ | # of elements (spanwise, wall-normal, streamwise adiabatic, heat flux) | Polynomial order | Total DoF |
|---|---|---|---|---|---|---|
| 5000 | 1.0 | 0.0 | 0 | 24,26,30,300 | 6 | 49 M |
| | | $5 \times 10^6$ | 0.2 | 24,30,30,300 | 8 | 134 M |
| | | $10^7$ | 0.4 | 24,30,30,300 | 8 | 134 M |
| | 0.0169 (lead) | 0.0 | 0 | 24,26,30,300 | 6 | 49 M |
| | | $5 \times 10^6$ | 0.2 | 24,30,30,300 | 8 | 134 M |
| | | $10^7$ | 0.4 | 24,30,30,300 | 8 | 134 M |
| | 0.0048 (sodium) | 0.0 | 0 | 24,26,30,300 | 6 | 49 M |
| | | $5 \times 10^6$ | 0.2 | 24,30,30,300 | 8 | 134 M |
| | | $10^7$ | 0.4 | 24,30,30,300 | 8 | 134 M |

**Table V. Catalog of data collected from presented DNS cases.**

| | | |
|---|---|---|
| Time averaged statistics (Whole domain) | Averaged velocity, temperature, and pressure | $<u^*>, <v^*>, <w^*>, <T^*>, <p^*>$ |
| | Spatial derivatives of averaged velocity and temperature | $\frac{\partial <u^*>}{\partial x_i}, \frac{\partial <v^*>}{\partial x_i}, \frac{\partial <w^*>}{\partial x_i}, \frac{\partial <T^*>}{\partial x_i}$ |
| | Root-mean-squared fluctuation (or variance if squared) | $u^{*\prime}_{rms}, v^{*\prime}_{rms}, w^{*\prime}_{rms}, T^{*\prime}_{rms}$ |
| | (Off-diagonal) Reynolds stress components | $<u^{*\prime}v^{*\prime}>, <v^{*\prime}w^{*\prime}>, <u^{*\prime}w^{*\prime}>$ |
| | Turbulent heat flux (THF) | $<u^{*\prime}T^{*\prime}>, <v^{*\prime}T^{*\prime}>, <w^{*\prime}T^{*\prime}>$ |
| | Turbulent kinetic energy (TKE)/Reynolds stresses budgets | $P_{ij}, \epsilon_{ij}, T_{ij}, \Pi_{ij}$ |
| | THF budgets | $P_{iT}, \epsilon_{iT}, T_{iT}, \phi_{iT}$ |
| Virtual probe-based (pointwise) | Time series of velocity and temperature | $u^*(t), v^*(t), w^*(t), T^*(t)$ |

**Table VI. Specification of data collection capability benchmark case.**

| | |
|---|---|
| $Re_\tau$ | 150 |
| $Pr$ | 0.71 |
| Geometric dimension $(L_x, L_y, L_z)$ (Spanwise, wall-normal, streamwise) | $(2\pi\delta, 2\delta, 10\delta)$ |
| Number of spectral elements $(N_x, N_y, N_z)$ (Spanwise, wall-normal, streamwise) | (20,15,13) |
| Statistics collection period | 5 domain flow through |

Figure 2 shows comparison of time-averaged statistics of benchmark case with the reference data set. The obtained results from benchmark case agrees well with the reference data.

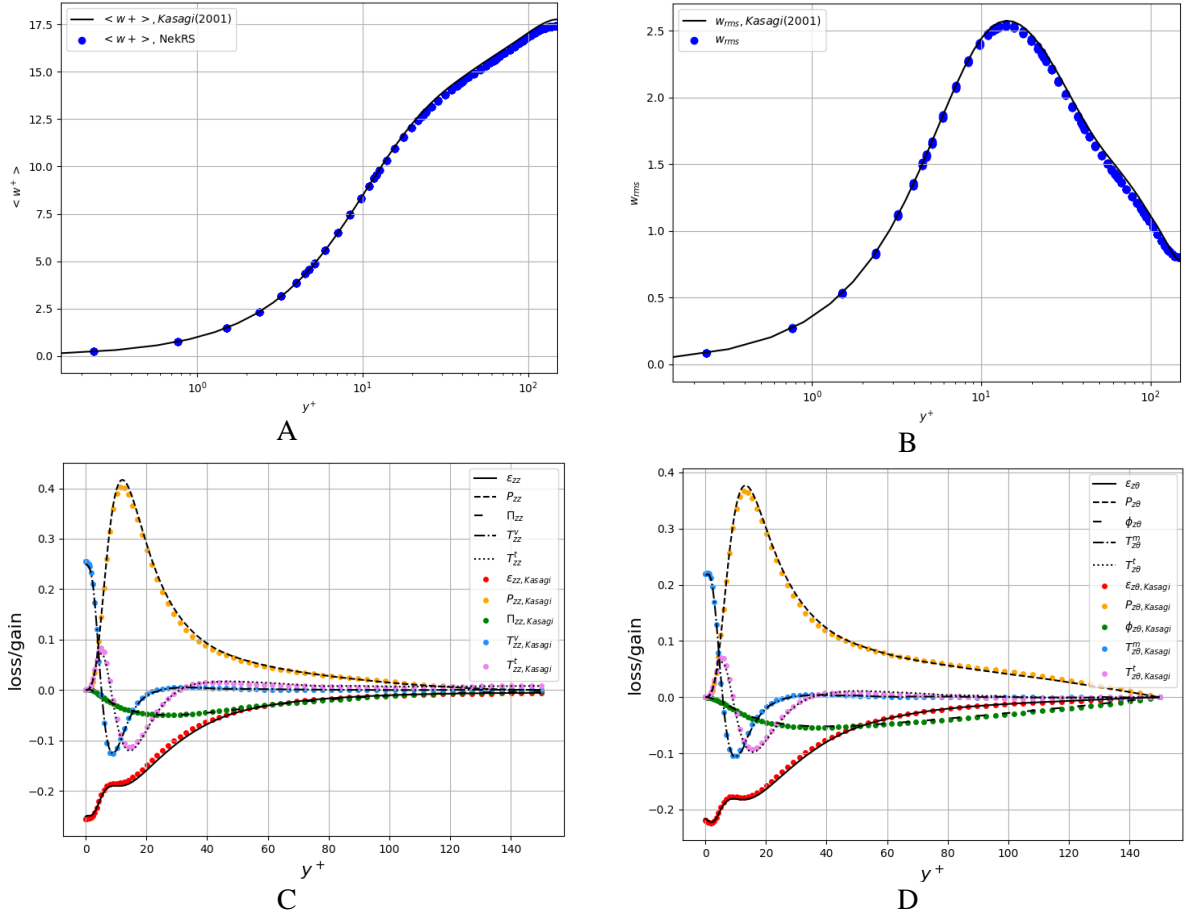

**Figure 2. Comparison of time-averaged statistics of benchmark case with reference data set [17].
A & B: mean-streamwise velocity and RMS fluctuation, C & D: TKE and THF budget terms.**

## 3. RESULTS AND DISCUSSION

### 3.1 Time-averaged Statistics

#### 3.1.1 Effect of buoyancy on mean flow statistics

Figure 3 shows mean and root-mean-squared streamwise velocity and temperature of presented cases. For the mean streamwise velocity, buoyancy effect rendered altered velocity boundary layer: wall shear near the buoyancy-aided/opposed (cooled/heated) wall is increased/reduced compared to the forced convection. Such effect is more obvious for liquid metal due to thicker thermal boundary layer. On the other hand, temperature profile of liquid metals is only slightly altered by buoyancy because conduction remains dominating heat transfer mode, whereas turbulence mixing effect leads to flattened temperature profile for unitary Prandtl mixed convection cases. For the RMS of streamwise velocity, flow opposed buoyant plume leads to monotonic enhancement of turbulence intensity at heated wall for all three presented $Pr$, and acceleration effect leads to reduction/reduction-then-increase of turbulence intensity at cooled wall for unitary/low $Pr$ cases. Also, compared to forced convection, significant enhancement of turbulence intensity is observed at channel centerline ($y = 0.0$) due to large scale buoyant plume rising from the heated wall.

Figure 4 shows profile of Reynolds shear stress, and THF on streamwise and wall-normal direction. Compared to forced convection, Reynolds shear stress peaking amplitude near heated wall showed significant increase and expansion into cooled wall due to the buoyant plume rising from the wall. Noticeable suppression of streamwise THF of unitary Prandtl cases is observed. For the liquid metal cases, though conduction remains a dominating heat transfer mode, buoyancy effect rendered increase in streamwise THF. On the other hand, wall normal THF enhancement is identified from all three Prandtl numbers with increasing $Ri$, showing the effect of buoyancy-induced turbulence mixing on turbulent heat transfer mode.

Figure 5 shows the production, dissipation, and molecular diffusion of TKE and streamwise THF. Compared to the forced convection, enhancement of TKE production near wall and channel center is observed in liquid metal cases, indicating effect of thermal plume interacting with the flow. Also, slight negative production is observed near cooled wall (buoyancy-aided side). This is due to the strong acceleration and flow laminarization effect due to the buoyancy. For unitary Prandtl cases, suppression of production is seen near the cooled wall.

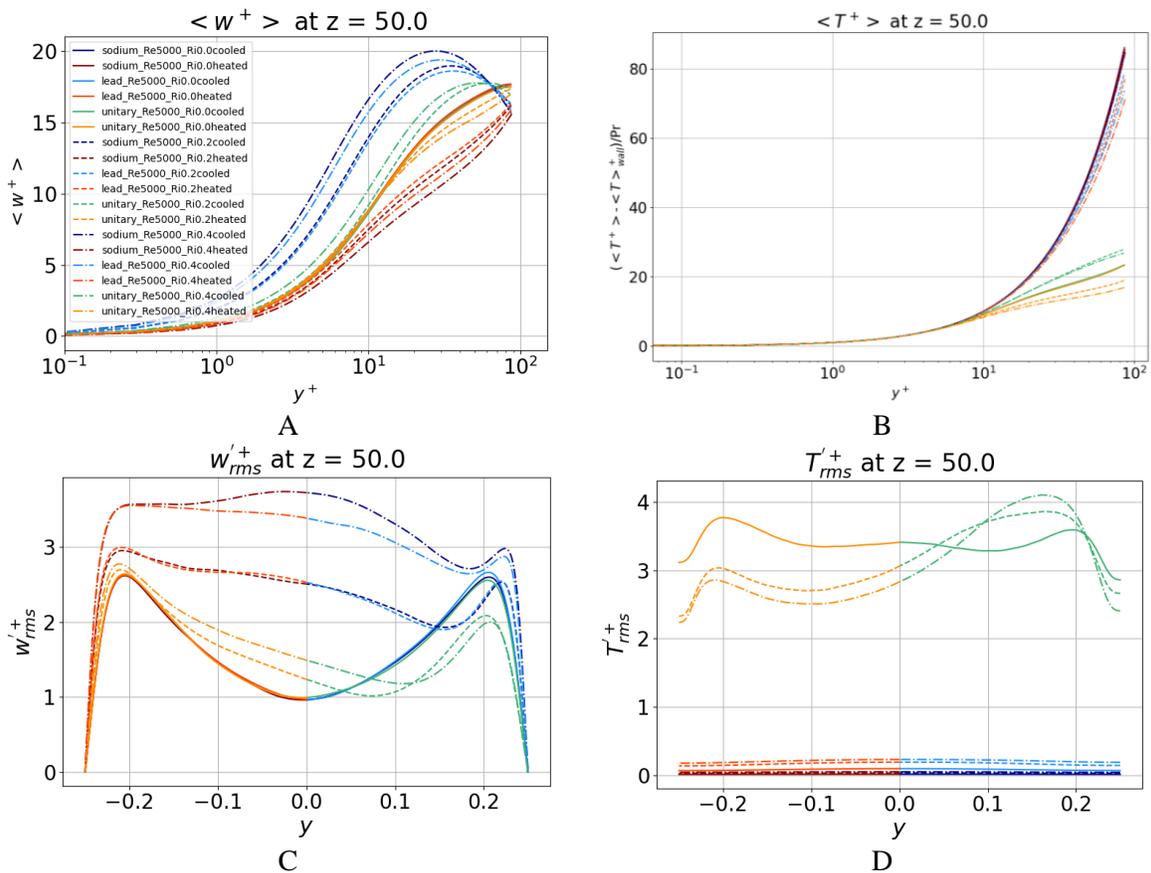

**Figure 3. Profile of A: mean streamwise velocity, B: mean temperature difference to the wall, C: RMS of streamwise velocity and D: RMS of temperature. Linear scale is adopted for x-axis of C and D for comparison between heated ($y = -0.25$) and cooled ($y = 0.25$) walls.**

### 3.1.2  Turbulent viscosity, diffusivity, and Prandtl number

One of the goals of the IRP NEAMS project is to develop data-driven Reynolds-averaged Navier-Stokes (RANS) turbulence models. Though, it is recognized that eddy viscosity and gradient diffusion hypotheses are inherently deficient for the considered flows [18], turbulent viscosity $\nu_t$, diffusivity $\alpha_t$, and Prandtl

$Pr_t = \nu_t/\alpha_t$ number are extracted since this may provide useful information for practitioners (simplest equations are employed).

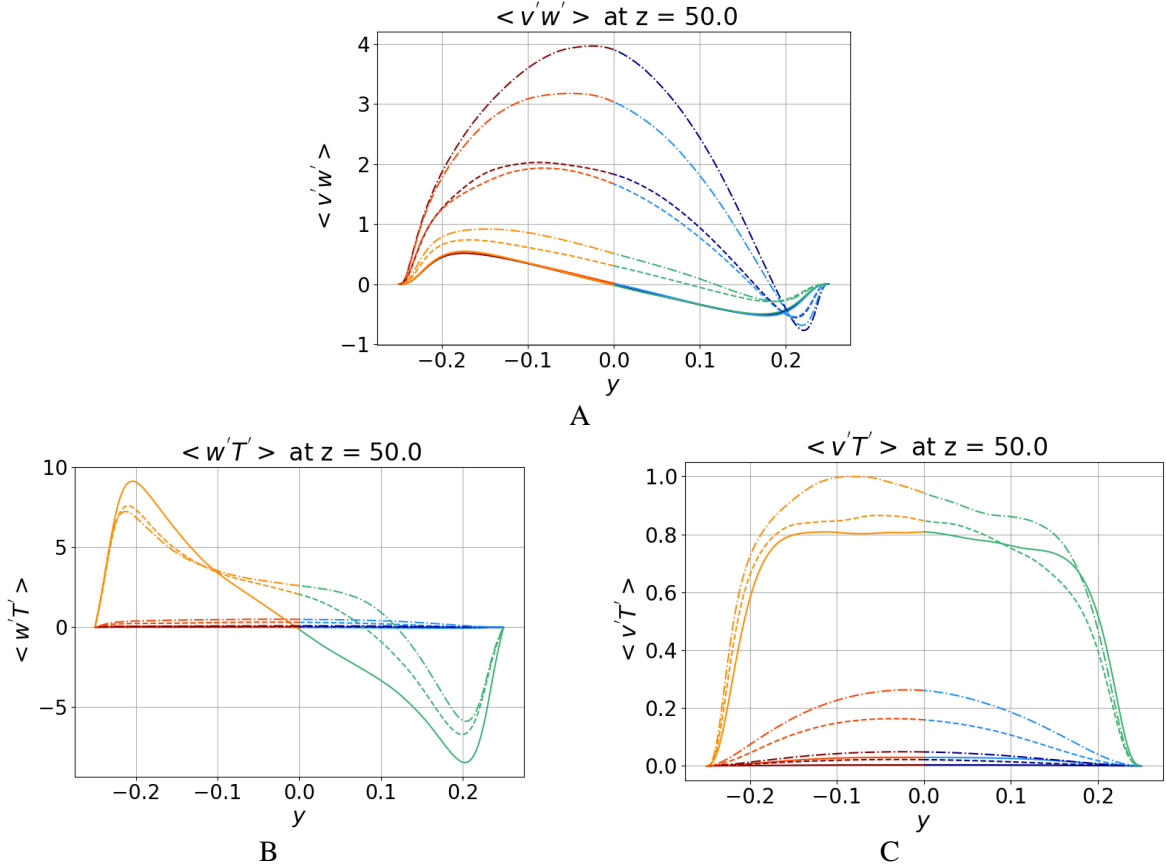

**Figure 4. Profile of A: Reynolds shear stress, B: streamwise THF, C: wall-normal THF. Note that linear scale is used in x-axis of the plot. See Figure 3A for legends.**

We note that when $Ri_q \neq 0$ negative or too large values of $\nu_t$ and $\alpha_t$ are inevitable. To get rid of such values, $\nu_t$ is linearly interpolated in between for 11 minimal values of $(\partial\langle v^*\rangle/\partial z + \partial\langle w^*\rangle/\partial y)$ at each $y$. For $\alpha_t$ the values higher than 99.7 percentile are cut off. For both $\nu_t$ and $\alpha_t$ negative values were nulled. Table VII summarizes ranges of the observed values (minimal value is assumed to be 0) within the heated region ($5 < z < 55$) and averaged $Pr_t$ for developing and fully developed regions. From Table VII one can see that the qualitative behavior is consistent for all considered fluids. As expected, when momentum and energy equations are decoupled ($Ri_q = 0$), $\nu_t$ is analogous for all fluids. For $Ri_q \neq 0$, $\nu_t$ is analogous for lead and sodium. Mean $Pr_t$ is in range 0.87-0.93 for $Pr = 1$, 1.6-3.3 for lead, and 3.4-9.4 for sodium depending on the conditions.

Figure 6 shows fields and profiles of the extracted quantities for lead with $Ri_q = 0.2$ in the developing region ($5 < z < 14$). One can see the difficulties in the modeling using the Boussinesq linear eddy viscosity hypothesis for such flows: there is a sharp discontinuity at $y \approx 0.15$ caused by the stratification (the discontinuity is slightly corrected by the linear interpolation). Such behavior greatly undermines the usefulness of $\nu_t$ in the future model development activities. There is also noticeable noisiness in $\nu_t$ that can be decreased by accumulating more temporal statistics and spatial filtering.

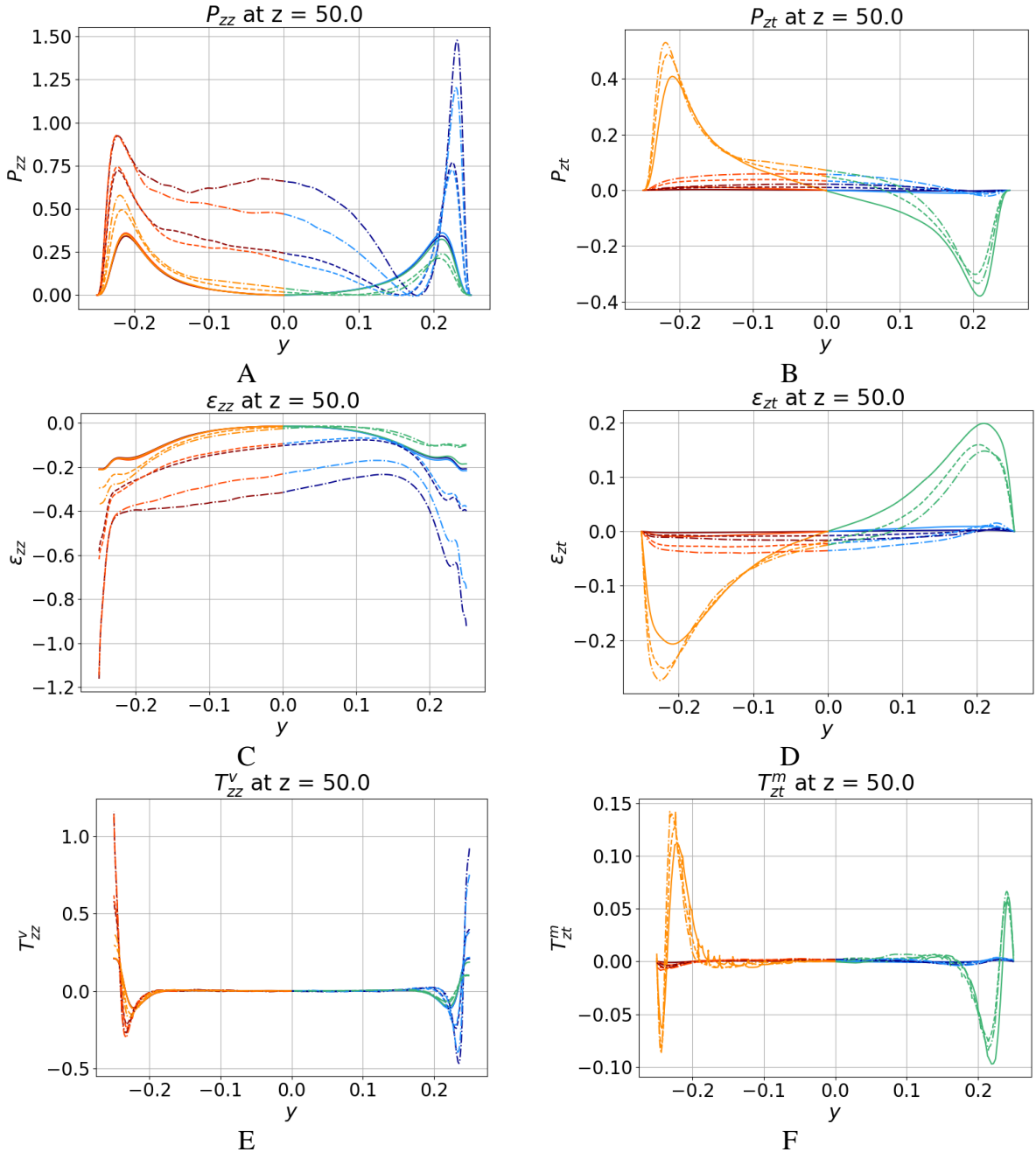

**Figure 5.** Profile of TKE and THF production (A, B), dissipation (C, D) and molecular diffusion (E, F). See Figure 3 for legends.

### 3.1.3. Nusselt number at fully developed region

Figure 7 shows the comparison of the averaged Nusselt number from $z \in [40,50]$ of presented cases. For unitary Prandtl cases, enhancement/impairment of convective heat transfer is observed at buoyancy opposed/aided side, which agrees with the trend suggested by Cotton and Jackson [3]. However, such trend is reversed in liquid metal cases, where enhancement/impairment of convective heat transfer is observed at buoyancy-aided/opposed side. For the presented lead cases, heat transfer impairment is not seen, but may happen at $Bo$ around 1.0 by extrapolation. Nusselt number correlation for three Prandtl numbers studied will be formed after more data sets are obtained.

Table VII. Ranges of observed $\nu_t$, $\alpha_t$, and $Pr_t$.

| $Re$ | $Pr$ | $Ri_q$ | $\nu_t$ (max) | $\alpha_t$ (max) | $Pr_t$ (max) | $Pr_t$ (mean) developing | $Pr_t$ (mean) fully developed |
|---|---|---|---|---|---|---|---|
| 5000 | 1.0 | 0 | 0.0018 | 0.0019 | 1.3 | 0.93 | 0.92 |
| | | 0.2 | 0.0028 | 0.0024 | 1.5 | 0.92 | 0.89 |
| | | 0.4 | 0.0035 | 0.0031 | 1.7 | 0.90 | 0.87 |
| | 0.0169 (lead) | 0 | 0.0018 | 0.00039 | 5.8 | 3.3 | 3.2 |
| | | 0.2 | 0.0067 | 0.0025 | 6.4 | 2.6 | 1.8 |
| | | 0.4 | 0.0103 | 0.0045 | 9.8 | 2.5 | 1.6 |
| | 0.0048 (sodium) | 0 | 0.0017 | 0.00013 | 14 | 9.3 | 9.4 |
| | | 0.2 | 0.0068 | 0.00102 | 22 | 6.6 | 4.6 |
| | | 0.4 | 0.0102 | 0.0021 | 25 | 5.4 | 3.4 |

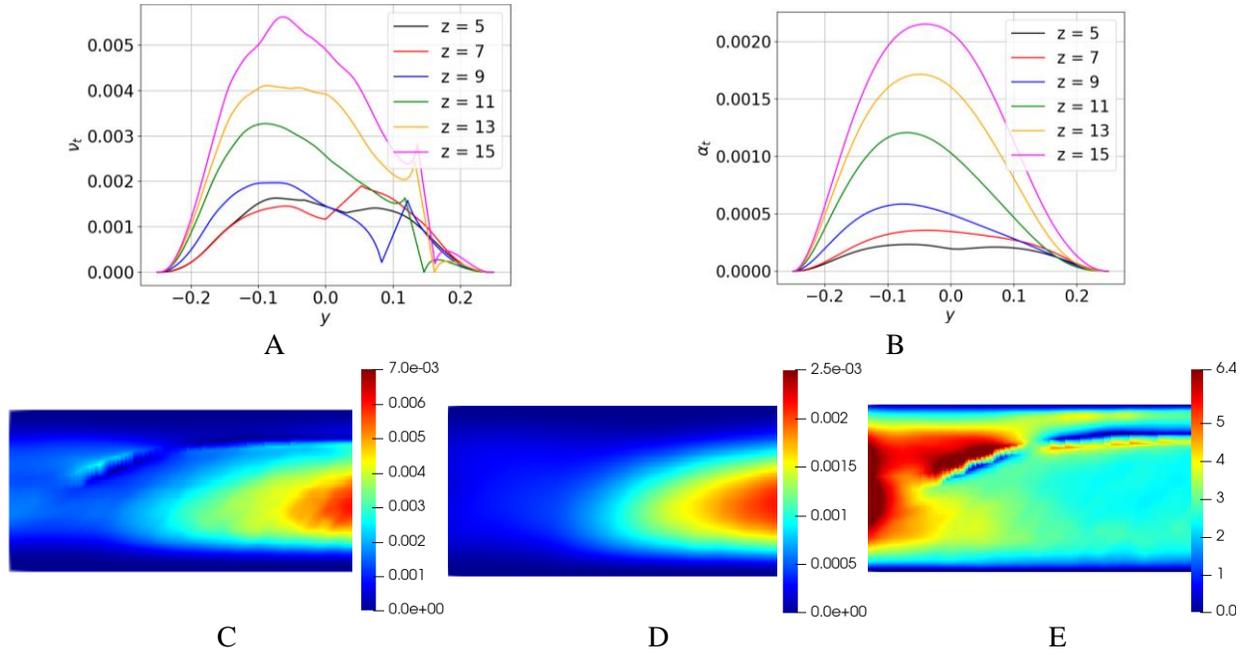

Figure 6. $\nu_t$, $\alpha_t$, and $Pr_t$ in the developing region for lead $Ri_q = 0.2$.

## 3.2 Instantaneous Time Signal and Frequency Analysis

To facilitate analysis on instantaneous behavior of mixed convection, time signals of velocity, and temperature of presented $Ri_q = 0.2$ cases are collected from various locations to facilitate analysis on instantaneous flow and heat transfer behaviors. Note that $z = 15$ in this case is in transition region whereas flow at $z = 50$ is fully developed. Figure 9 and the power spectral density (PSD) of streamwise velocity and temperature time signal near channel center and both walls and at both transition ($z = 15$) and fully developed ($z = 50$) region. In all presented PSDs, time series of length 80000 is employed. For velocity signal at the transition region, higher power density is observed below $100\ Hz$ near heated wall, suggesting buoyancy-induced enhancement in turbulence intensity. For the temperature signal, lower energy content over $3 - 30$ Hz is observed compared to velocity PSD due to the interplay between high thermal conductivity of sodium and buoyant production of turbulence. At the transition region, peaking of power density at around $10\ Hz$ appeared near the heated wall and channel centerline, suggesting possible large-scale temperature structure in this region. To identify possible coherent structure, wavelet transform using real-valued Morlet wavelet is applied to the fluctuation of the temperature signal. Results are shown in Figure 8. The signals from three probes at the same axial position are highly correlated due to the heat conduction remains the dominating mode of heat transfer for the selected case, and buoyant plume is developing from the heated wall, causing perturbation to the temperature field. Such coherence can also be seen from the wavelet spectrogram in Figure 8: similarity is observed at frequency range 30 to $10^2$ Hz, but with noticeable difference in the low-frequency range (~1 to 30 $Hz$). Further studies such as proper orthogonal decomposition will be applied to visualize the energy-dominating flow mode in this region.

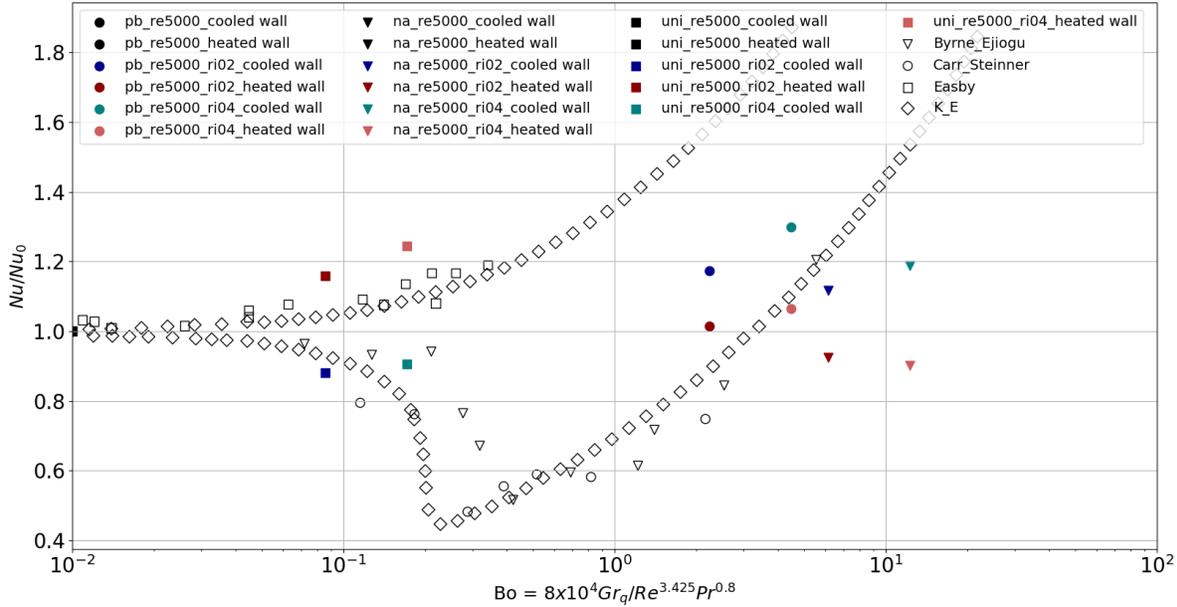

Figure 7. Comparison of Nusselt number at fully developed region with existing data sets [2].

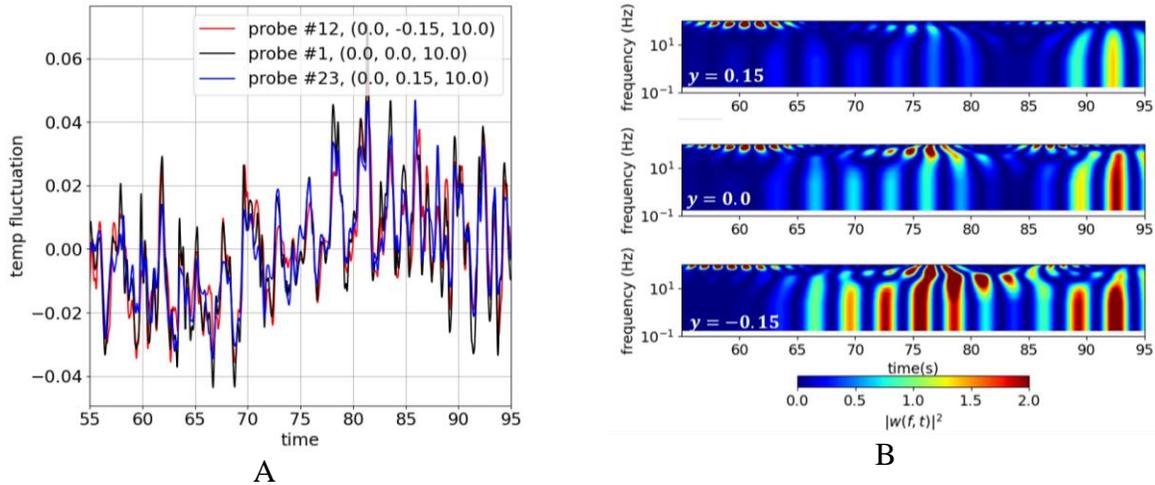

Figure 8 Fluctuation of temperature (A) and corresponding wavelet spectrogram (B) from probes placed at $z = 15$.

## 4    CONCLUSIONS

In this work, DNS is carried out to study the vertical mixed convection in downcomer representing canonical flows, with the flow conditions driven by the input of the industrial project partners. Analysis of fully developed flow statistics as well as time-dependent behavior are carried out based on the data collection workflow established in this work.

From the time-averaged statistics at the fully developed region, buoyancy altered the velocity boundary layer for liquid metal cases. Buoyancy also promoted/suppressed turbulence intensity near heated wall in all $Pr$ and in channel centerline for liquid metal cases. Negative TKE production can also be observed for sodium near cooled wall due to local flow laminarization. Effect of buoyancy also reflected on the heat transfer behavior: convective heat transfer is enhanced/impaired at buoyancy-opposed/aided side for unitary Prandtl cases and the trend is reversed for liquid sodium and lead. Wall-normal turbulent heat flux is enhanced in all presented cases with increasing strength of buoyancy ($Ri$). From time series analysis of sodium, $Re = 5000, Ri_q = 0.2$ case, large time-scale characteristics is observed and urges further study of flow in this region.

Broader data coverage in $Pr - Re - Pr$ parameter space will be provided in the future work. Besides different local flow physics, Buoyancy effect on Nusselt number can be correlated for different fluids of interest, which is crucial for engineering designs. In addition, physics at the transition region is not yet fully understood. Analysis techniques such as proper orthogonal decomposition will be conducted to probe the local flow behavior.

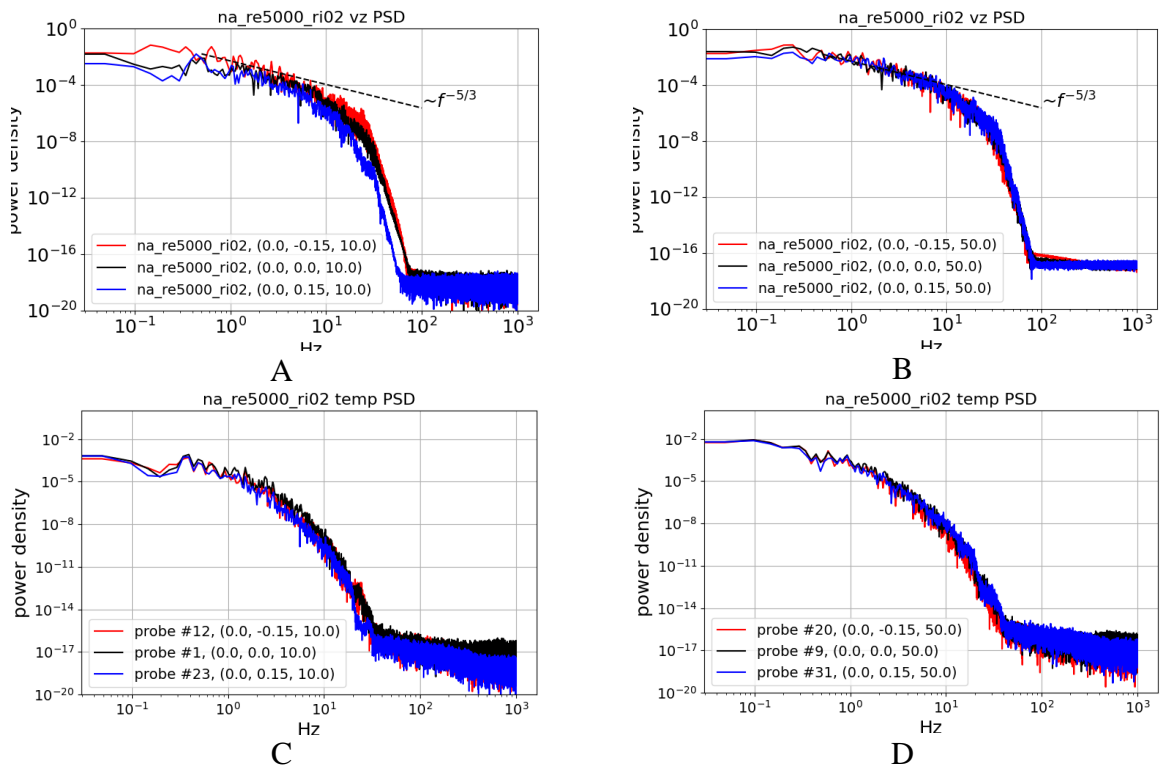

**Figure 9. PSD of streamwise velocity (A, B) and temperature (C, D) time signal of Na, $Re = 5000, Ri = 0.2$ case. $z = 15$ and $50$ correspond to the transition and fully developed region. Blue, black and red are used for time signal extracted from near cooled wall, channel centerline and near heated wall, respectively.**

**ACKNOWLEDGMENTS**

This work was funded by a U.S. Department of Energy Integrated Research Project entitled "Center of Excellence for Thermal-Fluids Applications in Nuclear Energy: Establishing the knowledgebase for thermal-hydraulic multiscale simulation to accelerate the deployment of advanced reactors -IRP-NEAMS-1.1: Thermal-Fluids Applications in Nuclear Energy. The computational resources are provided by the Oak Ridge Leadership Computing Facility at the Oak Ridge National Laboratory, which is supported by the Office of Science of the U.S. Department of Energy under Contract No. DE-AC05-00OR22725.

# NOMENCLATURE

**Nomenclature**

$c_p$: specific heat capacity
$\boldsymbol{g}$: gravitational acceleration vector
$u$: spanwise velocity
$v$: wall-normal velocity
$w$: streamwise velocity
$\boldsymbol{u}$: full 3D velocity vector
$T$: temperature
$T_0$: reference temperature
$T_{ij}, T_{i\theta}$: turbulent diffusion (TKE,THF budgets)
$t$: time
$P$: pressure
$P_{ij}, P_{i\theta}$: production (TKE, THF budgets)
$D_h$: hydraulic diameter ($D_h = 2L_y = 4\delta$)
$q''_{wall}$: wall heat flux

**Dimensionless groups**

$Re = \frac{|\boldsymbol{u}|D_h}{\nu}$: Reynolds number
$Gr_q$: Modified Grashof number
$Ri_q = \frac{Gr_q}{Re^2}$: Modified Richardson number
$Pr = \frac{\nu}{\alpha}$: Prandtl number
$Pr_t = \frac{\nu_t}{\alpha_t}$: turbulent Prandtl number
$Pe = RePr$: Peclet number
$Bo$: Buoyant number

**Greek alphabet**

$\alpha$: thermal diffusivity
$\alpha_t = -\langle v^{*\prime}T^{*\prime}\rangle/\frac{\partial\langle T^*\rangle}{\partial y}$: turbulent thermal diffusivity
$\beta$: rate of thermal expansion
$\delta$: half of wall spacing $\delta = \frac{L_y}{2}$
$\epsilon_{ij}, \epsilon_{i\theta}$: dissipation (TKE, THF budgets)
$\eta$: Kolmogorov length scale
$\lambda$: thermal conductivity
$\mu$: dynamic viscosity
$\nu$: kinematic viscosity
$\nu_t = -\langle v^{*\prime}w^{*\prime}\rangle/\frac{\partial\langle v^*\rangle}{\partial z} + \frac{\partial\langle w^*\rangle}{\partial y}$: eddy viscosity
$\phi_{i\theta}$: temp.-pressure grad. correlation (THF budgets)
$\rho$: density
$\tau_\eta$: Kolmogorov time scale
$\Pi_{ij}$: pressure-strain correlation (TKE budgets)

**Superscript, subscript, and operators**

$X_{ref}$: reference scales
$X^*$: dimensionless
$X^+$: $X^*$ normalized by viscous quantities
$X'$: fluctuation component, subtracted by mean
$<>$: time-averaging and spatial averaging
$X_{rms}$: root-mean-squared